\documentclass[12pt,epsf]{article}
\usepackage{graphicx}
\begin{document}


\def\a{\alpha}
\def\b{\beta}
\def\c{\varepsilon}
\def\d{\delta}
\def\e{\epsilon}
\def\f{\phi}
\def\g{\gamma}
\def\h{\theta}
\def\k{\kappa}
\def\l{\lambda}
\def\m{\mu}
\def\n{\nu}
\def\p{\psi}
\def\q{\partial}
\def\r{\rho}
\def\s{\sigma}
\def\t{\tau}
\def\u{\upsilon}
\def\v{\B}
\def\w{\omega}
\def\x{\xi}
\def\y{\eta}
\def\z{\zeta}
\def\D{\Delta}
\def\G{\Gamma}
\def\H{\Theta}
\def\L{\Lambda}
\def\F{\Phi}
\def\P{\Psi}
\def\S{\Sigma}

\def\o{\over}
\newcommand{\gsim}{ \mathop{}_{\textstyle \sim}^{\textstyle >} }
\newcommand{\lsim}{ \mathop{}_{\textstyle \sim}^{\textstyle <} }
\newcommand{\vev}[1]{ \left\langle {#1} \right\rangle }
\newcommand{\bra}[1]{ \langle {#1} | }
\newcommand{\ket}[1]{ | {#1} \rangle }
\newcommand{\EV}{ {\rm eV} }
\newcommand{\KEV}{ {\rm keV} }
\newcommand{\MEV}{ {\rm MeV} }
\newcommand{\GEV}{ {\rm GeV} }
\newcommand{\TEV}{ {\rm TeV} }
\def\diag{\mathop{\rm diag}\nolimits}
\def\Spin{\mathop{\rm Spin}}
\def\SO{\mathop{\rm SO}}
\def\O{\mathop{\rm O}}
\def\SU{\mathop{\rm SU}}
\def\U{\mathop{\rm U}}
\def\Sp{\mathop{\rm Sp}}
\def\SL{\mathop{\rm SL}}
\def\tr{\mathop{\rm tr}}

\def\IJMP{Int.~J.~Mod.~Phys. }
\def\MPL{Mod.~Phys.~Lett. }
\def\NP{Nucl.~Phys. }
\def\PL{Phys.~Lett. }
\def\PR{Phys.~Rev. }
\def\PRL{Phys.~Rev.~Lett. }
\def\PTP{Prog.~Theor.~Phys. }
\def\ZP{Z.~Phys. }
\newcommand{\bear}{\begin{array}}  \newcommand{\eear}{\end{array}}
\newcommand{\bea}{\begin{eqnarray}}  \newcommand{\eea}{\end{eqnarray}}
\newcommand{\beq}{\begin{equation}}  \newcommand{\eeq}{\end{equation}}
\newcommand{\bef}{\begin{figure}}  \newcommand{\eef}{\end{figure}}
\newcommand{\bec}{\begin{center}}  \newcommand{\eec}{\end{center}}
\newcommand{\non}{\nonumber}  \newcommand{\eqn}[1]{\beq {#1}\eeq}
\newcommand{\lmk}{\left(}  \newcommand{\rmk}{\right)}
\newcommand{\lkk}{\left[}  \newcommand{\rkk}{\right]}
\newcommand{\lhk}{\left \{ }  \newcommand{\rhk}{\right \} }
\newcommand{\del}{\partial}  \newcommand{\abs}[1]{\vert{#1}\vert}
\newcommand{\vect}[1]{\mbox{\boldmath${#1}$}}
\newcommand{\bib}{\bibitem} \newcommand{\new}{\newblock}
\newcommand{\la}{\left\langle} \newcommand{\ra}{\right\rangle}
\newcommand{\bfx}{{\bf x}} \newcommand{\bfk}{{\bf k}}
\newcommand{\gtilde} {~ \raisebox{-1ex}{$\stackrel{\textstyle >}{\sim}$} ~} 
\newcommand{\ltilde} {~ \raisebox{-1ex}{$\stackrel{\textstyle <}{\sim}$} ~}
\newcommand{\gtrsim}{ \mathop{}_{\textstyle \sim}^{\textstyle >} }
\newcommand{\lesssim}{ \mathop{}_{\textstyle \sim}^{\textstyle <} }
\newcommand{\ds}{\displaystyle}
\newcommand{\bi}{\bibitem}
\newcommand{\lar}{\leftarrow}
\newcommand{\rar}{\rightarrow}
\newcommand{\lrar}{\leftrightarrow}
\def\Frac#1#2{{\displaystyle\frac{#1}{#2}}}
\def\labelenumi{(\roman{enumi})}
\def\SEC#1{Sec.~\ref{#1}}
\def\FIG#1{Fig.~\ref{#1}}
\def\EQ#1{Eq.~(\ref{#1})}
\def\EQS#1{Eqs.~(\ref{#1})}

\newcommand{\fa}{F_{\cal A}}
\newcommand{\A}{{\cal A}}
\newcommand{\B}{{\cal B}}

\baselineskip 0.7cm

\begin{titlepage}

\begin{flushright}

\hfill DESY 05-261\\
\hfill hep-ph/0512296\\
\hfill February, 2006\\
\end{flushright}

\vskip 1.35cm
\begin{center}
{\large \bf
Unification of Dark Energy and Dark Matter
}
\vskip 1.2cm
Fuminobu Takahashi$^{1,2}$ and T. T. Yanagida$^{1,3}$
\vskip 0.4cm

${}^1${\it Deutsches Elektronen Synchrotron DESY, Notkestrasse 85,\\
22607 Hamburg, Germany}\\
${}^2${\it Institute for Cosmic Ray Research,
     University of Tokyo, \\Chiba 277-8582, Japan}\\
${}^3${\it Department of Physics, University of Tokyo,\\
     Tokyo 113-0033, Japan}

\vskip 1.5cm

\abstract{
 We propose a scenario in which dark energy and dark matter are
described in a unified manner.  The ultralight pseudo-Nambu-Goldstone (pNG) 
boson, $\A$, naturally explains the observed
magnitude of dark energy, while the bosonic supersymmetry partner of the pNG boson, $\B$,
can be a dominant component of dark matter. 
The decay of $\B$ into a pair of electron and positron 
may explain the 511 keV $\gamma$ ray from the Galactic Center.
 }
\end{center}
\end{titlepage}

\setcounter{page}{2}

\section{Introduction}

The cosmological constant problem is a long-standing 
problem. The recent
astrophysical observation~\cite{Riess:1998cb} 
has established that the expansion of the
present universe is accelerating,  indicating 
the existence of dark energy.
It is not excluded logically that the cosmological constant is 
tuned exactly zero in the true vacuum by a yet unknown mechanism, 
although the anthropic explanation of the observed small vacuum 
energy, $\Lambda_{\rm cos}^4 \simeq (2\times 10^{-3} {\rm eV})^4$, is quite 
natural~\cite{Weinberg:1987dv}. If it is the case, the present cosmological constant 
$\Lambda_{\rm cos}^4$ should be a potential energy carried by an extremely light boson called quintessence~\cite{Ratra:1987rm}, and the magnitude of $\Lambda_{\rm cos}$
may be directly linked to relevant energy scales beyond the standard model.

Some years ago,  it was
pointed out that a pseudo-Nambu-Goldstone (pNG) boson ${\cal A}$
coupled to left-handed-lepton currents plays a role of the quintessence~\cite{Nomura:2000yk}. 
In fact, one-loop diagrams induce the pNG boson coupling to the weak 
SU(2)$_L$ gauge fields. And SU(2)$_L$ instantons generate 
a potential for the pNG boson as
\begin{equation}
       V\simeq Ce^{-2\pi/\alpha_2(F_{\cal A})}m_{3/2}^3F_{\cal A}
(1-{\rm cos}({\cal A}/F_{\cal A})),   
\end{equation}
where $C$ is a constant of order 1 and $F_{\cal A}$ the decay constant
of the pNG boson ${\cal A}$. Provided $F_{\cal A}\simeq M_{PL}\simeq
2.4\times 10^{18}$ GeV, the observation on $\Lambda_{\rm cos}$ suggests
the gravitino mass is $m_{3/2}\simeq {\cal O}(10)$ MeV. Here, we have
used $\alpha_2(M_{PL})\simeq 1/23$. To be more precise we will consider a bit
larger scale of the symmetry breaking, $F_{\cal A}\simeq 4\pi M_{PL}$.
This is because the quintessence ${\cal A}$ has already started to roll
down to the potential minimal, otherwise. We do not, however, take 
too large value
for the decay constant $F_{\cal A}$, since the born unitality of
graviton and graviton scattering processes is violated at such energy
scale and our field-theory description becomes no longer valid. 

In this short note, we point out that the rather small gravitino mass 
leads to an interesting effect on cosmology. A scalar boson $\B$
which is a bosonic supersymmetry (SUSY) partner of ${\cal A}$ has a mass 
of order $m_{3/2} \simeq 10$ MeV.
Since the life time of this boson $\B$ is so long, the $\B$ density
easily overcloses the present universe. However, the energy density does
crucially depend on  the cosmological history. 
If late-time entropy production~\cite{Lyth:1995ka,Kawasaki:2004rx} occurred,  the $\B$ density
should have been diluted. In fact  it is known that a thermal
inflation~\cite{Lyth:1995ka} at the weak scale dilutes the energy density substantially,
and renders the $\B$ field to be a dominant component of the dark matter
in the present universe. Therefore, in our scenario,  the dark energy and dark matter 
are unified, in a sense that they are explained by the pNG boson $\A$ and its bosonic SUSY partner $\B$,
respectively. See Refs.~\cite{chaplygin,quintessence} for different approaches to unified description of
dark energy and dark matter. 

We would stress that  ${\cal A}$ and $\B$ have couplings with a pair
of electron and positron as
\begin{equation}
     -{\cal L} =  i m_e \frac{\A}{\sqrt{2} F_{\cal A}}   \,\bar e \gamma_5 e+
     		      m_e   \frac{\B}{\sqrt{2} F_{\cal A}} \, \bar e e.                      
\end{equation}
Thus, the dark matter $\B$ decays into a pair of electron and
positron. The life-time of  $\B$ is much longer than the age
of the present universe owing to the large scale of $F_{\cal A}$.
It is very encouraging that the decay of $\B$ into $e+\bar e$ may 
explain the $511$ keV $\gamma$ ray excess from the 
bulge of our galaxy observed by SPI/INTEGRAL~\cite{SPI}
(See also Refs.~\cite{DDM1,DDM2,Kawasaki:2005xj}).

In the next section, we provide a model which unifies dark energy and 
dark matter.
We will give discussions in the last section.

\section{Unification Model}

We denote the pNG chiral multiplet as $S$ whose bosonic component $s$
is consist of the pNG boson ${\cal A}$ and the scalar boson $\B$;
\begin{equation}
      s = \frac{\B + i{\cal A}}{\sqrt{2}}.
\end{equation}
The global U(1) symmetry is represented by a shift symmetry,
\begin{equation}
      S \rightarrow S + i \alpha \fa,
\end{equation}
where $\alpha$ is a real constant. We see that the K\"ahler potential
$K(S+S^{\dagger})$ is invariant under the symmetry. The K\"ahler potential is
written as
\begin{equation}
      K =       \fa^2 \left[\frac{1}{2}\left(S/ \fa+S^{\dagger} /\fa \right)^2 + 
     \kappa_3 \left(S/ \fa+S^{\dagger} /\fa\right)^3  +  \kappa_4 \left(S/ \fa+S^{\dagger} /\fa\right)^4
     +\cdots \right]
\end{equation}
Here, we have absorbed the linear term of $S+S^{\dagger}$ into the
definition of the field $S$.
We further impose that the left-handed lepton doublets $\ell_i~ (i=1\sim3)$
have the U(1) charge and they transform under the global symmetry as
\begin{equation}
      \ell_i\rightarrow e^{-i \alpha}\ell_i.
\end{equation}
Then, we have an invariant superpotential,
\begin{equation}
      W= f_i e^{S/F_{\cal A}} e_i\ell_iH_d,
\end{equation}
which leads to the following interaction in the electroweak-symmetry
breaking vacuum ($\la H_d^0 \ra \ne 0$);
\begin{equation}
\label{eq:ee-int}
      -{\cal L} = i m_e  \frac{ \A }{ \sqrt{2} F_{\cal A}} \,\bar e \gamma_5 e +
                      m_e   \frac{ \B}{ \sqrt{2} F_{\cal A}}\, \bar e e                     
			+\cdots.
\end{equation}
The above global symmetry has a SU(2)$_L$ gauge anomalies and one-loop
diagrams of the internal left-handed leptons induce 
\begin{equation}
\label{}
      {\cal L}_{\rm kin} =  \int d^2 \theta \left[
      \frac{3}{32 \pi^2} \frac{S}{\fa} W^{a \alpha} W^a_\alpha \right]+ {\rm h.c.},
\end{equation}
where $W^a_\alpha$ is the gauge kinetic function of the weak
SU(2)$_L$ gauge multiplet, and $a=1\sim3$ and $\alpha = 1,2$ run over the SU(2) generators
and the components of spinors, respectively.  This gives the anomaly interaction of the pNG
boson ${\cal A}$ as
\begin{equation}
      {\cal L} = -\frac{3}{32 \pi^2} \frac{\cal A}{\sqrt{2}F_{\cal A}}F^a_{\mu\nu} \tilde F^{a \mu\nu}.
\end{equation}
The potential of ${\cal A}$ is generated by SU(2) instanton~\cite{Nomura:2000yk}:
\bea
V &\simeq&\Lambda_{\cal A}^4
\left(1-{\rm cos}({\cal A}/F_{\cal A})\right)
\eea
with 
\bea
\Lambda_{\cal A}^4 &\simeq& Ce^{-2\pi/\alpha_2(F_{\cal A})}m_{3/2}^3F_{\cal A},
\eea
where $C$ is a constant of order unity.
Here let us consider the dynamics of ${\cal A}$, assuming  
 $\Lambda_{\cal A} \sim \left(10^{-3} {\rm eV}\right)^4$. If $\fa \simeq M_{PL}$, the mass of $\A$
becomes roughly equal to the present Hubble parameter, which means that $\A$ has already
started rolling down to the potential minimum. To circumvent this we adopt slightly
larger value of $\fa \simeq 4 \pi M_{PL}$~\footnote{
The theory with such large cut-off has been proposed in 
another context~\cite{Ibe:2004mp}. In this scheme,
due to the large cut-off scale, inflation tends to predict almost scale invariant spectrum.
The little hierarchy between $M_{PL}$ and $\fa$ might be ascribed to the anthropic explanation~\cite{Ibe:2004mp}.
}. Interestingly enough,
such value of $\fa$ turns out to be favorable in the context of explaining 
511 keV $\gamma$ line, as we will see later.
The potential energy of ${\cal A}$  accounts for the
present dark energy, if  the gravitino mass $m_{3/2}$ is $O(10)$MeV~\footnote{
In Ref.~\cite{Nomura:2000yk}, the gravitino mass was fixed to the weak scale.
That is why the authors introduced flavor symmetry to suppress the potential.
In this paper, instead, we determine the gravitino mass from the magnitude of
the cosmological constant $\Lambda_{\rm cos}^4$.
}:
\bea
\Lambda_{\cal A}^4 &\simeq&C \left(1 \times 10^{-3} {\rm eV}\right)^4  \left(\frac{m_{3/2}}{15{\rm MeV}}\right)^3 
\left(\frac{\fa}{4 \pi M_{PL}}\right),
\eea
where  we have used $\alpha_2^{-1}(\fa) \simeq 
\alpha_2^{-1}(M_{PL}) -1/(2 \pi) \cdot \ln{(4 \pi)}$.

Now let us turn to $\B$, the bosonic SUSY partner of $\A$. 
The $\B$ field obtains a soft SUSY breaking mass, $m_\beta$,  which is of 
order $m_{3/2} = O(10)$MeV.
The position of $\B$ during inflation is generally displaced from the potential
minimum after inflation by $M_{PL}$, unless we impose additional symmetry~\footnote{
\label{ft:sym}
The initial displacement of $\B$ from the potential minimum can be
suppressed by imposing a symmetry. For instance,
let us take the following superpotential;
\beq
W = X (\Phi \bar{\Phi} -\mu^2),
\eeq
where $X(0)$, $\Phi(+1)$ and $\bar \Phi(-1)$ are scalar chiral superfields with the charges 
of the global $U(1)$ symmetry  shown in the parentheses, and $\mu$ is the breaking 
scale of the $U(1)$. In this model, $\B$ corresponds to the difference$\sim \Phi-\bar \Phi$,
and $\B$ is lifted by a soft SUSY breaking mass.
 Assuming a symmetry interchanging $\Phi$ with $\bar\Phi$, the potential of $\B$
has a minimum at the origin, $\B = 0$. If $\Phi$ and $\bar\Phi$ acquire 
Hubble-induced masses during inflation, $\B$ rolls down to the origin, which should coincide with the potential minimum after inflation.  Then the $\B$ density due to
the initial misalignment is exactly zero. Therefore it is possible to set
the energy density of $\B$
to the right amount of dark matter by introducing a tiny violation of the interchanging symmetry. 
Such violation may naturally arise from the fact $\Phi$ and $\bar \Phi$ interact with matter fields
in a different way due to $U(1)$ charge assignment.
 }. After inflation, $\B$ starts oscillating around the minimum when the Hubble parameter becomes 
comparable to its mass, $H \simeq m_\B$.
Since the life time of $\B$ is so long due to the large scale $\fa$, 
the energy density $\rho_\B$ may overclose the universe. However, 
the present value of $\rho_\B$ crucially depends on the thermal history of the
universe; if the $\B$ density is diluted by late-time entropy production, $\B$
can be a dominant component of dark matter. Here we estimate the requisite 
amount of entropy production. Before the late-time entropy production,
the primordial $\B$ density is
\beq
\label{eq:primordial_B}
\frac{\rho_\B}{s_i} \sim 10^7{\rm GeV} \left(\frac{m_{\B}}{10{\rm MeV}}\right)^{\frac{1}{2}}
 \left(\frac{\B_i}{M_{PL}}\right)^{2},
\eeq
where $\B_i$ denotes the initial amplitude of $\B$, $s_i$ the entropy density and
we have assumed the reheating is completed when $\B$ starts oscillating.
Thus, in order to render $\B$ to the dark matter, it must be off at least by
$10^{16}$. 

The thermal inflation is able to generate the needed entropy. According to 
Refs.~\cite{Hashiba:1997rp}, 
the thermal inflation is driven by the potential energy of the flaton, $\phi$,
with the potential
\beq
V(\phi) = V_0 + (T^2-m_0^2) \phi^2 + \frac{\phi^6}{M_{PL}^2},
\eeq
where $T$ denotes the cosmic temperature, and $V_0 \sim m_0^3 M_{PL}$.
The thermal inflation lasts between $T \sim m_0^{3/4} M_{PL}^{1/4}$ and $T \sim m_0$. 
After the thermal inflation,
the flaton oscillates around the minimum $\phi \sim \sqrt{m_0 M_{PL}}$ and 
decays, producing large entropy. 
The entropy dilution factor is 
\beq
\Delta_{TI} \equiv \frac{s_f}{s_i} \simeq 10^{20} \left(\frac{10{\rm MeV}}{T_{d}}\right),
\eeq
where $s_f$ is the entropy density after thermal inflation, $T_d$ is the decay 
temperature of the flaton~\footnote{For successful BBN, $T_d$ cannot be much smaller
than $10$MeV. See also Refs.~\cite{Kawasaki:1999na}.}. The primordial $\B$ density after thermal inflation is then
\beq
\label{eq:primordial_omegaB}
\Omega_\B^{\rm p} = O(10^{-3})  \left(\frac{m_{\B}}{10{\rm MeV}}\right)^{\frac{1}{2}}
 \left(\frac{\B_i}{M_{PL}}\right)^{2}  \left(\frac{T_d}{100{\rm MeV}}\right).
\eeq
On the other hand, the thermal
inflation itself displaces the potential minimum of $\B$ from that in the vacuum,
re-generating $\B$ density at the end of the thermal inflation. The $\B$ density
generated in this way is estimated as
\beq
\label{eq:TI_omegaB}
\Omega_\B^{\rm TI} \simeq O(0.1)  \left(\frac{10{\rm MeV}}{m_\B}\right)^{2}
 \left(\frac{m_0}{100{\rm GeV}}\right)^{3}  \left(\frac{T_d}{100{\rm MeV}}\right).
\eeq
The total $\B$ density is given by the sum of (\ref{eq:primordial_omegaB}) and (\ref{eq:TI_omegaB}).
Therefore $\B$ can constitute a dominant part of the dark matter.

The important feature of the $\B$ dark matter is that it decays into a pair
of electron and positron through the interaction (\ref{eq:ee-int}).
The decay rate is~\footnote{
$\B$ can decay into two $\A$s as well, if $\kappa_3$ is nonzero. 
Therefore it is model-dependent whether this decay channel becomes important or not.
In fact, the dominant decay
channel would be $\B \rightarrow \A+\A$ if $\kappa_3$ is order unity. 
However, even in this case, the life time of $\B$ is still much longer than the 
age of the present universe, therefore
the predicted 511keV $\gamma$ ray flux (\ref{eq:511}) remains valid, since it is determined
by only $\Gamma_{\B \rightarrow \bar e^+ e^-}$, not the total decay rate.}
\bea
\Gamma_{\B \rightarrow \bar e^+ e^-} &=& \frac{m_e^2 m_{\beta}}{16 \pi \fa^2},\non\\
&\simeq& \left( 10^{25}\, {\rm sec} \right)^{-1} 
\left(\frac{m_{\beta}}{10{\rm MeV}}\right)  
			\left(\frac{4 \pi M_{PL}}{\fa}\right) ^2.
			\eea
The produced positrons will annihilate mostly by forming positroniums~\cite{SPI}, 
a quarter of which produce 511keV line $\gamma$ ray. 
It is amazing that thus obtained decay rate explains the 511keV $\gamma$ ray from
the Galactic Center observed by SPI/INTEGRAL~\cite{SPI} as
pointed out in Refs.~\cite{DDM2,Kawasaki:2005xj}. The $\gamma$ ray flux is
estimated as
\beq
\label{eq:511}
\Phi_{511}^\B \sim 10^{-3} {\rm ph\,\,cm}^{-2} {\rm sec}^{-1} 
\left(\frac{10{\rm MeV}}{m_{\beta}}\right)
 \left(\frac{10^{25} \,{\rm sec}}{\Gamma_{\B \rightarrow \bar e^+ e^-}^{-1} }\right), 
\eeq
while the observed flux is~\cite{SPI}
\beq
\Phi^{\rm obs}_{511} = (1.05 \pm 0.06) \times 10^{-3}  {\rm ph\,\,cm}^{-2} {\rm sec}^{-1} .
\eeq
In deriving \EQ{eq:511}, we have assumed the following dark matter density
function~\cite{Navarro:2003ew} with $\alpha = 0.1$,
\beq
\rho_{\rm DM} (r) = \rho_0 \exp\left[-\frac{2}{\alpha}\left(\left(\frac{r}{r_0}\right)^\alpha-1
\right)\right]
\eeq
where $r_0 = 20 h^{-1}$kpc, $\rho_0$ is normalized so that $\rho_{\rm DM}(r=8.5{\rm kpc})=0.3{\rm GeV}/{\rm cm}^3$.

\section{Discussions}
In the previous section we have concentrated on the bosonic part of
the pNG chiral multiplet $S$; the imaginary component $\A$ explains the dark energy,
while the real one $\B$ becomes dark matter. Here let us consider
the fermionic component, $\tilde{s}$. Its mass is of order $m_{3/2}$ 
and the cut-off scale of the interactions with leptons is $\fa$.
Therefore the $\tilde{s}$-abundance is always smaller than the gravitino abundance.
In addition, the late-time entropy production dilutes the $\tilde{s}$ density,
so $\tilde{s}$ does not play any important role in the history of the universe.

Since the late time entropy production dilutes any pre-existing baryon asymmetry,
we need to either (i) generate large enough baryon asymmetry before entropy
production; or (ii) generate baryon asymmetry {\it after} entropy production.
The former does not seem to work since the requisite entropy dilution is very huge~\cite{Kasuya:2001tp}.
In the latter case, the promising baryogenesis is the Affleck-Dine leptogenesis 
after thermal inflation~\cite{Stewart:1996ai, Jeong:2004hy}. It should be noted,
however, that such constraints on the baryogenesis mechanism disappear if the
$\B$ density is suppressed by a symmetry discussed in footnote \ref{ft:sym}.

Very recently, Ref.~\cite{Beacom:2005hg} has appeared and put a stringent bound
on the injection energies of  positrons $\lesssim 3{\rm MeV}$, by using the observed
Galactic gamma-ray data. This bound corresponds to
$m_\B \lesssim 6{\rm MeV}$ in our scenario, which can be easily satisfied
 since  $m_\B$ does not necessarily coincides with $m_{3/2} \sim 10{\rm MeV}$, 
 and may be a few times  smaller.
  
Our model has another observational implication; $\B$ decays into two photons
through one-loop diagrams, producing the line gamma rays with energy $m_\B/2$, 
as pointed out in Ref.~\cite{Kawasaki:2005xj}.  The line gamma ray flux is 
so small that it is below the bound from present data~\cite{Kawasaki:2005xj}.
Therefore future observation on the gamma ray background may be able to support or refute
our scenario.

\section*{Acknowledgments}
F.T. is grateful to Motoi Endo for useful discussions, and 
W. Buchm\"uller for  comments. We thank  J. Beacom for comments.
F.T.  would like to thank the Japan Society for Promotion of 
Science for financial support. The work of T.T.Y. has been supported in part by
a Humboldt Research Award.  T.T.Y. thanks theory group at DESY for the hospitality
during the stay.


\end{document}